# A multi-scale approach to microstructure-sensitive thermal fatigue in solder joints

Yilun Xu[1]*, Jingwei Xian[1]*, Stoyan Stoyanov[2], Chris Bailey[2], Richard J. Coyle[3], Christopher M. Gourlay[1] and Fionn P. E. Dunne[1]

[1]Department of Materials, Imperial College, London SW7 2AZ, UK
[2]Computational Mechanics and Reliability Group, University of Greenwich, London SE10 9LS, UK
[3] Nokia Bell Labs, Murray Hill, New Jersey, USA

**Abstract**

This paper presents a multi-scale modelling approach to investigate the underpinning mechanisms of microstructure-sensitive damage of single crystal Sn-3Ag-0.5Cu (wt%, SAC305) solder joints of a Ball Grid Array (BGA) board assembly subject to thermal cycling. The multi-scale scheme couples board-scale modelling at the continuum macro-scale and individual solder modelling at the crystal micro-scale. Systematic studies of tin crystal orientation and its role in fatigue damage have been compared to experimental observations. Crystallographic orientation is examined with respect to damage development, providing evidence-based optimal solder microstructural design for in-service thermomechanical fatigue.

[1] Corresponding author: j.xian@imperial.ac.uk; yilun.xu@imperial.ac.uk

# 1 Introduction

This paper utilizes multi-scale modelling to investigate the crystallographic orientation effect on the microstructure-sensitive deformation and damage within SAC305 solder joints subject to thermal cycling. The multi-scale model couples a continuum model at the BGA board scale and a dislocation-based crystal plasticity model at the individual solder scale, and the numerical predictions are compared to the experimental characterization. Various typical crystallographic orientations of the Sn phase are then assessed using the proposed multi-scale model to reveal the underpinning mechanisms for the microstructure-sensitive damage development within solders, which would in turn facilitate the microstructure optimization of solders and the improvement of the reliability for electronic components attached to printed circuit boards.

Thermal fatigue of solder interconnects in electronic components attached to printed circuit boards is a major failure mechanism. The thermomechanical damage induced in a solder joint is caused by operational temperature cycling leading to thermal stress due to mismatch between thermal expansion coefficients (CTE) of the assembly materials (Yao et al., 2017) and any additional mechanical constraints. Therefore, the durability of solder interconnects to thermomechanical fatigue is an important design objective for solder alloys (Coyle et al., 2015). Among the Pb-free solders, Sn-Ag-Cu (SAC) alloys have become the most widely used (Coyle et al., 2015) thanks to their combination of solderability and performance (Miller et al., 1994; Suganuma, 2001). There has been extensive research on characterising and understanding the relationships between the solder microstructure and the thermomechanical performance of SAC305 and other lead-free solder alloys (e.g. (Bieler et al., 2012;

Coyle et al., 2015)). The anisotropy of elasticity and thermal expansion with the Sn phase contained in SAC305 was identified as potentially the most important factor for the damage degradation of this solder alloy (Matin et al., 2005; Matin et al., 2007). The effect of its c-axis (i.e. [001] direction) orientation with respect to the BGA substrate is also considered to be a key factor affecting the damage development in solders (Bieler et al., 2012), with a c-axis perpendicular to the substrate observed to be the most fatigue resistant orientation (Bieler et al., 2008).

Optical microscopy and digital image correlation (DIC) characterization was performed to reveal the role of strain and its inhomogeneity on solder damage under thermal cycling (Park et al., 2007). In-situ electron backscattered diffraction (EBSD) analysis was performed to evidence slip activation and microstructural deformation as a precursor to failure of SAC305 single crystal in thermal cycling (Gu et al., 2020). However, due to the complexity of both the microstructure of SAC305 alloy and the complex loading, the underpinning mechanisms for thermomechanical fatigue failure of solder joints in electronic component assemblies remain incompletely understood. The combined effect of the solder joint location in the grid array for BGA components and the solder microstructure (and notably the combined effects of crystal orientation and the length-scale of intermetallics (Xu et al., 2021a) on the resulting thermal stresses, slip activation, and in turn the solder damage has received relatively little attention. The identification and understanding of the microstructure features which predominate in the fatigue performance remain to be established in SAC305 solder joints.

Various BGA board-scale models of the package assembly (Stoyanov et al., 2020) have been developed and reported which assess statistically or via physics-of-failure simulation approaches the deformation regimes, thermal fatigue damage, and lifetime of solder joints in BGA assemblies but without recognising any damage development at the microstructure scale. However, fatigue damage originates at the microstructural length scale, initiating with slip activation and dislocation pile-up with the consequent stress concentrations at the intragranular scale (Gustafson et al., 2020; Jiang and Dasgupta, 2021). In relation to research on the damage mechanisms of individual solder joints, several continuum models (Darbandi et al., 2014; Jiang and Dasgupta, 2021; Zamiri et al., 2009) were developed for SAC alloys, but intermetallic size and morphology effects (along with crystallographic anisotropy) were not considered, and recently it has been observed that they strongly affect the creep behaviour of SAC305 alloy (Xu et al., 2021a) using a particle-hardening model (Li et al., 2020). Previous studies (Maleki et al., 2011, 2013) did not fully include the crystallographic elasto-plastic anisotropy for the Sn phase, which has been extensively considered as a major driver for damage development within solders (Bai and Chen, 2009; Gong et al., 2007; Zhou et al., 2015) or the scale coupling to BGA board (Mukherjee et al., 2016). Therefore, a major challenge yet to be addressed is connecting the thermo-mechanical response and behaviour of the BGA array assembly of solder joints to the individual solder joint response driven by its anisotropic microstructure attributes.

In this study, a multi-scale methodology is proposed that couples modelling of the BGA assembly (component-board level) using a macro-scale continuum formulation and the modelling of the individual solder joint behaviour using non-linear, temperature-dependent, microstructure-governed crystal plasticity techniques. The proposed multi-scale approach differs from other published analysis frameworks as the models at the two scales are fully coupled by passing thermomechanical boundary conditions from the solder board scale down to individual solder beads, in order to explicitly capture how Sn crystal orientation affects reliability in single grain joints with similar IMC particle size. It is the first application to Sn-based solder alloys of a length-scale dependent, stored energy density driver of fatigue crack nucleation, which has provided mechanistic understanding of microcrack nucleation in other alloys (Chen et al., 2018a; Guan et al., 2017; Prastiti et al., 2020). The link between solder crystallographic orientation, local misorientation (lattice curvature) and thermomechanical fatigue cracks in solders has been established, and a mechanistic explanation presented for the relationship between solder grain c-axis orientation and corresponding fatigue life.

The innovation in this paper is the combination of modelling approaches at two length-scales, revealing how factors at both length-scales interact to determine the optimum Sn-phase orientation for fatigue performance in thermal cycling. Specifically, the approach allows us to conclude that the optimum microstructure depends on the c-axis crystallographic orientation relative to both the plane on the substrate and to the neutral point vector.

The next section details the experimental procedure and observations of solder cyclic thermomechanical loading in a BGA component. A multi-scale numerical modelling approach for the analysis of solder joint damage at the joint level is then presented together with a comprehensive parametric study of the role of $\beta$-Sn orientation in single grain solder joints. Findings and recommendations about optimizing the microstructure of single crystal solder joints subject to thermal cycling are proposed.

## 2   Experiments
### 2.1   BGA Component and Load history

The BGA component is a 7x7 mm 84CTBGA thin core package that features a 12x12 ball grid array at pitch size 0.5 mm (Figure 1(a)) soldered with 300μm SAC305 balls and matching SAC305 paste. Details of this test vehicle including the board and package attributes, attachment process and reflow soldering are given in (Coyle et al., 2016). The whole BGA board was subjected to the cyclic temperature loading detailed in Figure 1(b), representing a typical in-service thermal load for consumer/handheld electronic devices (Coyle et al., 2016). The thermal cycle is defined with temperature extremes 0°C and 100°C, ramp rate of 10°C/minute and dwell times (of 10 minutes) at the low and high temperature extremes. Thus, the total duration of one temperature cycle is 40 minutes. Because the rate of temperature change is low, the quasi-static loading (no thermal shock) may be assumed to give rise to uniform (but time-varying) temperature for the entire assembly domain (Sidhu and Chawla, 2008).

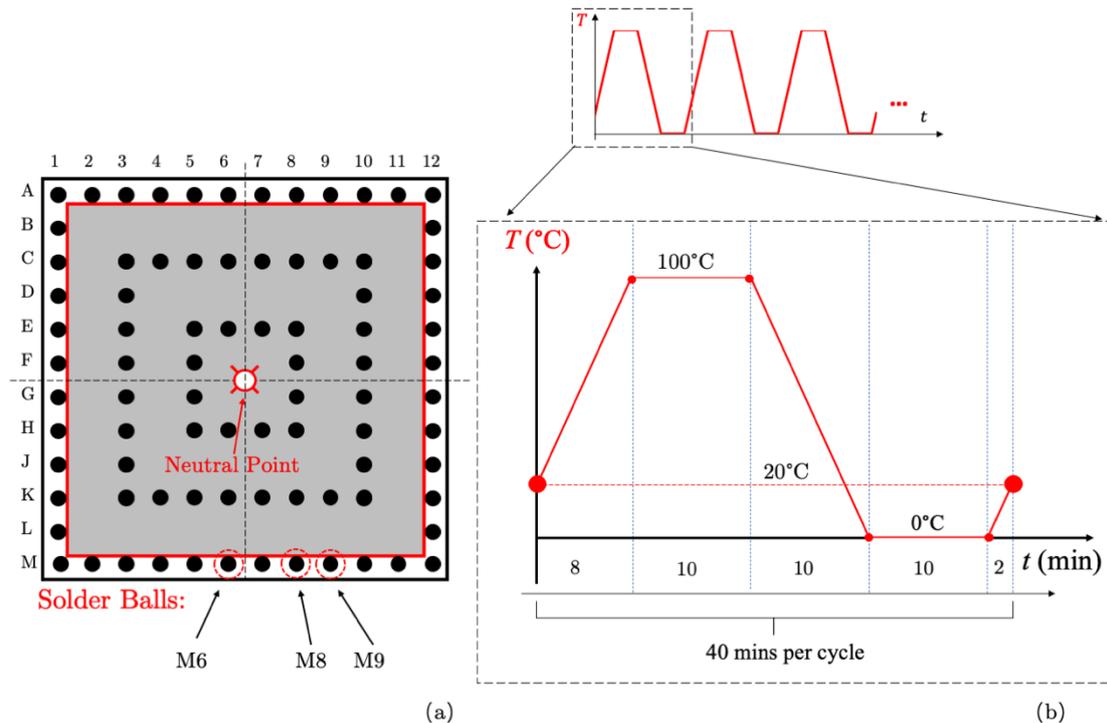

Figure 1. (a) schematic diagram of BGA solder ball array. The solder joints investigated in this study are indicated by red circles i.e. M6, M8 and M9, (b) the temperature cycles applied to the BGA assembly.

## 2.2 Sample preparation and characterization

Two samples of the 84CTBGA package ((Figure 1(a)) were subjected to 7580 thermal cycles with the thermal profile in Figure 1. After thermal cycling, the package/solder/PCB cross-sections were mounted in Struers VersoCit acrylic cold mounting resin and prepared using standard metallography procedures to study row M. The cross-sections were imaged in a Zeiss Auriga field emission gun scanning electron microscope (FEG-SEM) with crystallographic orientations collected by a Bruker e-FlashHR EBSD detector. Among all 24 joints in the M rows from the two packages, only three were single β-Sn grain joints and were at positions M6 in package 1 and M8-M9 in package 2.

Figure 2(a) shows SEM images of each joint with an insert showing a typical backscattered electron (BSE) image of fine eutectic $Cu_6Sn_5$ and $Ag_3Sn$ particles.

Eutectic particle size measurements were conducted on BSE images from these joints using an image segmentation method (Xian et al., 2021). The intermetallic particle size distribution was studied in M8 and M9 and was highly similar with mean, standard deviation, median of 0.52, 0.28, 0.44 µm for M8 and 0.54, 0.29, 0.45 µm for M9. Therefore, the IMC size variation is deemed negligible for the joints in this study.

Figure 2(b) shows EBSD phase maps. Most eutectic $Ag_3Sn$ particles (green) were too small to be detected with the EBSD settings used, however some larger $Ag_3Sn$ particles near the top regions can be seen in solder joints M8 and M9. Orange $Cu_6Sn_5$ particles in the EBSD phase maps appear larger than their actual sizes measured by BSE images, which is due to the slightly protruding IMC particles on the polished surface and consequent shading effects during EBSD mapping.

Figure 2(c) shows the IPF-Y maps of the β-Sn phase. It can be seen that post-cycling β-Sn orientations are mostly uniform across each joint except for the top regions near the package side where damage is localised. Therefore, it is reasonable to assume the as-solidified β-Sn orientations are largely unchanged during thermal cycling near the central area of the joints. The mean β-Sn reference orientations were obtained by averaging within selected rectangular boxes (Figure 2(c)) and are plotted as unit cell wireframes beneath each joint in Figure 2(c). Using MATLAB™ 9.2 (Mathworks, USA) with the MTEX 5.1 toolbox (Bachmann et al., 2010), the misorientation distributions in each joint were calculated with respect to the mean β-Sn reference orientations as shown in Figure 2(d). It can be seen that thermal fatigue damage (*i.e.* misorientation distribution) is localised near the package side (top) interface of the joints, and recrystalized b-Sn grains are also localised in this region. It can also be seen that the damage (both the extent of misorientation and the area of recrystallised grains) varies in the joints, increasing from M6 to M8 to M9. The strong correlation between the misorientation developed and experimentally observed crack nucleation sites within solders (top interface) is addressed in a later section.

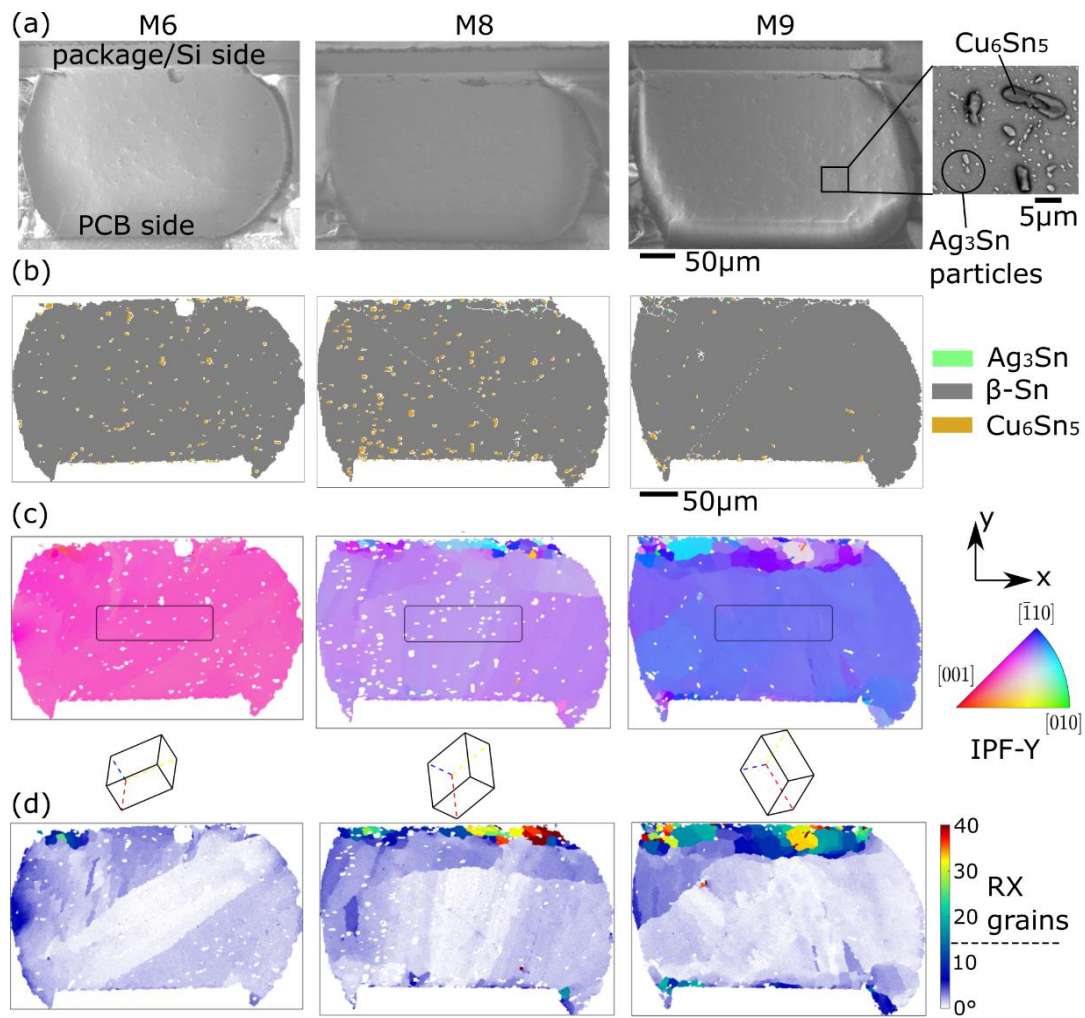

Figure 2. EBSD analysis of three single-crystal joints in row M of 84CTBGA packages after 7580 thermal cycles from 0/100°C. (a) SEM images with insert showing Ag$_3$Sn and Cu$_6$Sn$_5$ particles. (b) EBSD phase maps. (c) EBSD IPF-Y maps each with a unit cell wireframe of the mean orientation taken from the central black reference boxes. (d) Misorientation maps with reference to the mean orientations in (c); recrystallised (RX) grains have MO >~15°.

# 3  The multi-scale modelling approach

The multi-scale modelling scheme to investigate the solder joint damage resulting from thermal fatigue loading comprises the following three key aspects:

(1) **Continuum Model:** A global level continuum finite element model is applied to simulate the thermo-mechanical response of the entire BGA assembly under the thermal cyclic load. Solders at this stage are modelled with (isotropic) non-linear viscoplastic behaviour and are taken to be homogeneous.

(2) **Coupling between models:** Dataflow from the component-level model to the microstructure model: the top and bottom solder boundary three-dimensional cyclic displacement results at the single joint boundaries are extracted and provided as loading boundary conditions for the explicit microstructural solder crystal plasticity modelling.

(3) **Microstructural Model**: Explicit model representation of single solders (including crystal orientation) incorporating anisotropic elasticity and thermal expansion as well as anisotropic dislocation-based crystal slip, with joint-interface displacement, and temperature loading conditions, passed down from the continuum model, allowing investigation of microstructural solder effects.

An overview of the multi-scale modelling strategy is outlined in Figure 3.  In this paper we discuss and report only the unidirectional (i.e. uncoupled) modelling methodology which passes the displacement and temperature information, in the form of boundary conditions, from the continuum model to the crystal plasticity model. Ongoing research is addressing the fully coupled approach where microstructure-sensitive parameters identified with the crystal plasticity model will be fed back and inform the material modelling of solder at the continuum macro-scale. Previous work has addressed coupling the length scales within individual solders (Jiang et al., 2022; Mukherjee et al., 2016), but the link between the scales of the BGA board and the individual solder has not yet been fully established. This prevented the application of the numerical model to investigate the solder location effect. In the proposed multi-scale modelling framework, we transfer the information from the BGA board scale through the continuum model to the individual solder scale by virtue of ensuring the displacement compatibility along interfaces. In addition, the micro-scale deformation obtained at the solder scale is compared to the experimental observation of mis-orientation that may eventually drive recrystallization (Chen et al., 2020) and/or failure.

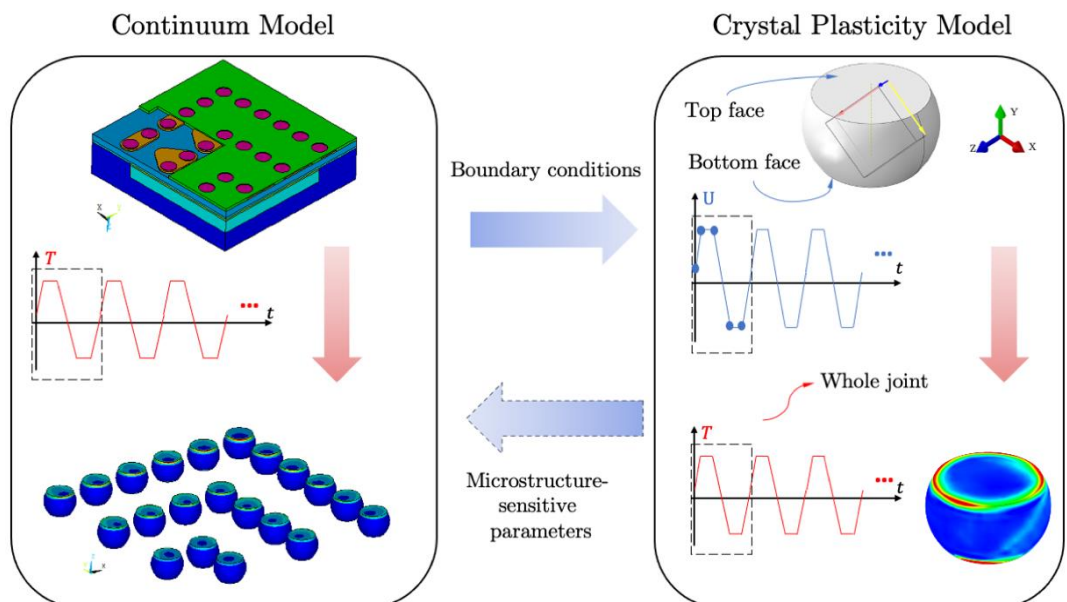

Figure 3. Overview of the multi-scale modelling approach applied to investigate the damage developed within solder joints subject to the thermal fatigue

The modelling methodology is applied to the package detailed in section 2, with three different microstructures being analysed for the solder positions of interest. The BGA package (continuum) modelling approach is outlined in Section3.1, and a concise summary of the solder microstructural (crystal plasticity – CP) modelling for the Sn phase and its material properties is provided in Section3.2.

## 3.1  Continuum modelling at BGA board-scale

Micro-section images of the BGA assemblies were used to measure and confirm the package and solder joint dimensions (*e.g.* representative stand-off height) when building the model. A full three-dimensional macro-scale finite element model of the BGA assembly was developed to simulate the thermo-mechanical response of the structure under the cycling thermal load. The model development is undertaken using the finite element simulation software ANSYS APDL. Because of the existing symmetry in the package architecture and the ball layout, only a quarter section of the whole assembly is represented. The internal structure of the BGA component is explicitly modelled. It includes the silicon die, substrate with outer solder mask layers, contact pads, the pad metallisation, and the epoxy moulding compound (EMC) encapsulation of the full package. All solder joints within the modelled quarter section of the assembly are represented. The layered PCB is modelled as a homogenised structure, although the outer layers of solder mask and the PCB copper pads with respective traces coming out are explicit.

The finite element mesh model is based on the ANSYS element technology for 3-D modelling of solids (SOLID185) and has a full range of non-liner capabilities such as plasticity and creep. The element is defined by eight nodes having three degrees of freedom at each node (translations in the nodal x, y, and z directions). Mesh convergence checks have been undertaken by assessing three different mesh densities with 30% mesh size increments, for the 3D CAD model of the BGA assembly. The nodal displacements associated with the spatial locations of the minimum and maximum displacement in the joints during temperature cycling were analysed and confirmed that full converge across the three mesh models was achieved.

The macro-level BGA assembly model is detailed in Figure 4.

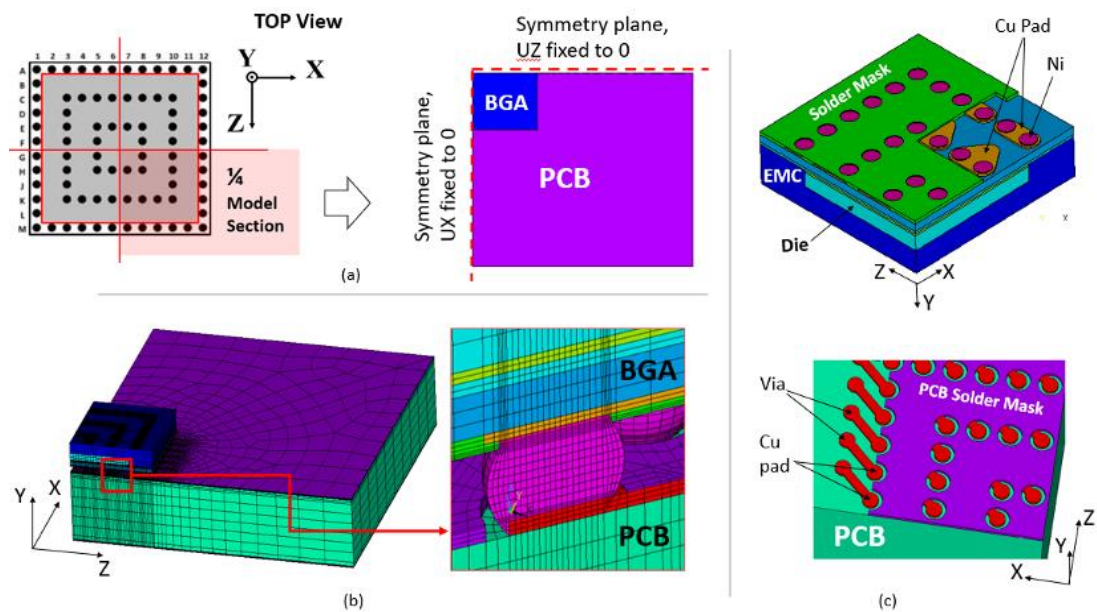

Figure 4. Global level finite element model of the BGA assembly: (a) illustration of the existing symmetry of the assembly and the quarter section of the domain captured with the finite element model, (b) mesh of the finite element model and a zoom-in view of the mesh density at the level of the solder joints, and (c) spatial features of the BGA and the PCB included in the model.

The stress-free temperature of the whole package is set to 150°C in the assembly macro-model. This is assumed to correspond to the processing temperature at which the package is encapsulated (the epoxy moulding compound is cured) and the multi-layer PCB stack is laminated (the pre-preg epoxy resin is cured). At this processing temperature, the BGA and PCB are assumed stress free as the epoxy resins are not hardened, and it is only during the subsequent cooling down from the curing temperature of the respective fabrication process when residual stresses start to develop and become locked in the structure. For the SAC305 solder, the stress-free temperature is different, and it is set to 217°C, the liquidus temperature of the solder during reflow.

The non-linear transient simulations for predicting the thermo-mechanical responses of the whole BGA assembly, and in particular the solder joint responses, under the cyclic temperature loading are undertaken using ANSYS simulation software. Five temperature cycles were simulated to provide a stabilised prediction for the solder deformation excursion over one temperature cycle as well as a global-level model prediction for the thermal fatigue damage of the solder joints based on inelastic strain energy density results (Che and Pang, 2013; Darveaux, 2002).

The required material properties of the BGA assembly on the PCB are gathered through measurements and from technical datasheets, and are partly complemented with data from public sources (Chen et al., 2018b; Lall and Wei, 2016; Motalab et al., 2013). The assembly materials, except the solder and copper, are modelled with elastic behaviour. Their Young's modulus and coefficient of thermal expansion (CTE) are defined with temperature dependant values. The PCB and BGA substrate are layered composites and hence the analysis deploys their respective orthotropic properties. Copper (BGA and PCB pads) is modelled with elastic-plastic behaviour. The material properties used in the analysis are summarised in Table 1.

Table 1. Material properties used in the BGA finite element analysis.

| Material | Young's Modulus \| Shear Modulus (GPa) | Coefficient of Thermal Expansion, CTE (ppm/°C) | Poisson's ratio |
|---|---|---|---|

| | | | | |
|---|---|---|---|---|
| BGA | Silicon | 169 | 1.88 @ -40°C<br>2.61 @ 25°C<br>2.80 @ 50°C<br>3.25 @ 125°C | 0.28 |
| | Mould Compound | 23.0 @ -40°C<br>18.0 @ 120°C<br>2.3 @ 140°C<br>2.3 @ 150°C | 15.0 @ -40°C<br>15.0 @ 120 C<br>43.0 @ 140 C<br>43.0 @ 150 C | 0.3 |
| | BGA Package Substrate | 17.9 (x) \| 8.1 (xz)<br>7.8 (y) \| 2.8 (zy)<br>17.9 (z) \| 2.8 (xy) | 12.4 (x)<br>57.0 (y)<br>12.4 (z) | 0.11 (xz)<br>0.39 (zy)<br>0.39 (xy) |
| | Die Attach | 6.8 | 52.0 | 0.35 |
| | Nickel (pad metallisation) | 205 | 13.1 | 0.31 |
| | Copper (BGA pads) | 110<br>Yield Stress: 0.172<br>Tangent Modulus: 1.034 | 15.3 @ -40°C<br>16.4 @ 25°C<br>16.7 @ 50°C<br>17.3 @ 125°C | 0.34 |
| | Solder mask | 3.1 | 30.0 | 0.3 |
| PCB | Composite multi-layer stack | 22 (x) \| 10.4 (xz)<br>10 (y) \| 3.6 (zy)<br>24 (z) \| 3.6 (xy) | 16.1 (x)<br>44.0 (y)<br>13.5 (z) | 0.11 (xz)<br>0.39 (zy)<br>0.39 (xy) |
| | Solder mask | 3.1 | 30.0 | 0.3 |
| Solder Joints | SAC305 (homogenised) | 57.4 @ -40°C<br>46.5 @ 25°C<br>38.2 @ 75°C<br>30.0 @ 125°C | 22.5 | 0.4 |

*Note: Orthotropic property data: x and z in-plane , y out-of-plane*

In the macro-model of the BGA assembly, the SAC305 solder joints are assumed to be homogeneous and to exhibit visco-plastic behaviour that is modelled using the Anand inelastic strain rate material model (Anand, 1985). The Anand model consists of a flow equation governing the inelastic strain rate,

$$\frac{d\varepsilon_p}{dt} = A\left[\sinh\left(\xi\frac{\sigma}{s}\right)\right]^{\frac{1}{m}} exp\left(-\frac{Q}{RT}\right) \tag{1}$$

and two evolution equations governing the internal state variable ($s$) and its saturation value ($s^*$):

$$\frac{ds}{dt} = \left\{h_0(|B|)^a \frac{B}{|B|}\right\} \frac{d\varepsilon_p}{dt}$$

$$s^* = \hat{s} \left\{\frac{\left(\frac{d\varepsilon_p}{dt}\right)}{A} exp\left(\frac{Q}{RT}\right)\right\}^n \tag{2}$$

where

$$a > 1 \quad B = 1 - \frac{s}{s^*} \tag{3}$$

and the complete nomenclature list for the parameters in the above equations and their respective units is provided in Table 2. The same table details the Anand model constants for SAC305 solder deployed in this analysis (Basit et al., 2015; Motalab et al., 2012; Motalab et al., 2013). The Anand model for the SAC305 solder alloy used in this study has been fully validated against experimental data and excellent agreement between the model and measured data has been reported (Motalab, 2012).

Table 2. Anand model constants for Sn3.0Ag0.5Cu.

| Model Parameter | Parameter Value | Description |
|---|---|---|
| $A$ (s$^{-1}$) | 3501 | Pre-exponential factor |
| $Q/R$ (1/K) | 9320 | Q: activation energy, R: universal gas constant |
| $\xi$ | 4.0 | Stress multiplier |
| $m$ | 0.25 | Strain rate sensitivity of stress |
| $\hat{s}$ (MPa) | 30.2 | Coefficient of deformation resistance saturation value |
| $n$ | 0.01 | Strain rate sensitivity of saturation (deformation resistance) value |
| $h_0$ (MPa) | 180,000 | Hardening/softening constant |
| $a$ | 1.78 | Strain rate sensitivity of hardening or softening |
| $s_0$ (MPa) | 21.0 | Initial value of deformation resistance |

### 3.2 Microstructural modelling at individual solder scale

The simulation at the individual solder scale takes the solder interface displacements passed from the board-scale modelling, as well as the temperature-time history, and imposes them onto a solder model where crystal plasticity finite element (CPFE) modelling is utilised.

### 3.2.1 Individual solder modelling

The displacement fields at both top and bottom interfaces of the solder joint are transferred from the continuum model results to the individual solder scale modelling, which is schematically shown in Figure 5. Five loading cycles are simulated in the board-level modelling allowing a steady state to be achieved. The stabilized cyclic displacement field obtained is then utilised for the individual solder CP model, and five consecutive cycles (reflecting the $6^{th}$ -$10^{th}$ BGA cycles) are modelled in the CP model. The displacements spatially distributed over the solder upper and lower interfaces are passed to the CP modelling node-wise, for five key time points (indicated by the solid circles) shown in the thermal cycle in Figure 5. The effects of the geometrical details near the upper and lower solder interfaces to the BGA board, such as the mask, and mould compound are carried through faithfully to the individual solder CP model by virtue of extracting continuum (BGA) model displacements at the cutting planes shown in Figure 5. The solder is discretised into 2,620 (element size $\approx$ 4μm) $2^{nd}$ order, reduced-integration, hexagonal-shaped elements. This approach significantly improves the efficiency of both solder geometry representation and the CPFE calculations without compromising the accuracy. The representative crystallographic orientations for the investigated solder single crystals are measured using EBSD and indicated using a unit cell shown in Figure 5. The material properties are obtained from a microstructure-based, multi-scale homogenization study, and the material properties are introduced in Section 3.2.3.

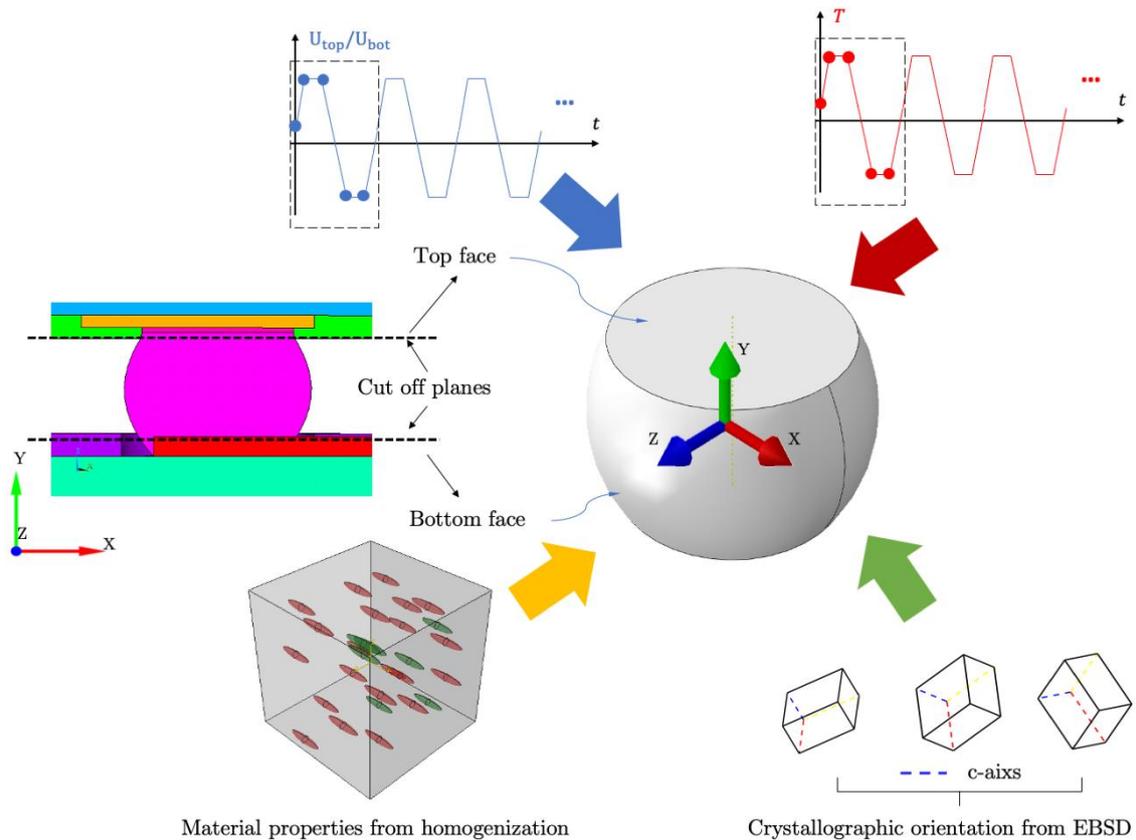

Figure 5. the crystal plasticity model for the individual solder joint. Its displacement boundary conditions are imposed on the top and bottom surfaces, which are passed from the BGA board-scale model. The temperature history is applied throughout the whole joint. The crystallographic orientation is taken as the one for Sn phase measured using EBSD, and the material properties are obtained from previous physics-based homogenization.

### 3.2.2 Crystal plasticity modeling

A strain-rate sensitive, temperature-dependent crystal plasticity finite element framework is adopted for modelling the SAC305 crystal deformation at the individual solder scale. Only a brief summary is provided here, and details can be found in other literature *e.g.* (Dunne et al., 2007).

At a given loading instant, the total deformation gradient $\boldsymbol{F}$ is decomposed into three terms, namely the elastic $\boldsymbol{F}^e$, plastic $\boldsymbol{F}^p$ and thermal $\boldsymbol{F}^\theta$ deformation gradients which give:

$$\boldsymbol{F} = \boldsymbol{F}^e \boldsymbol{F}^p \boldsymbol{F}^\theta \tag{4}$$

The rate of thermal deformation gradient $\dot{\boldsymbol{F}}^\theta$ is given by:

$$\dot{\boldsymbol{F}}^\theta = \dot{\theta}\boldsymbol{\alpha}_{th}\boldsymbol{F}^\theta \tag{5}$$

Where $\dot{\theta}$ is the rate of change of temperature, and $\boldsymbol{\alpha}_{th}$ is the anisotropic thermal expansion coefficient vector with respect to the global coordinate system.

The plastic strain rate $\dot{\boldsymbol{\varepsilon}}^p$ at any instant is given by the sum of projected shear strain rates $\dot{\gamma}^{(i)}$ along corresponding active slip systems, which gives:

$$\dot{\boldsymbol{\varepsilon}}^p = \text{sym}(\sum_{i=1}^{M} \dot{\gamma}^{(i)} \boldsymbol{s}^{(i)} \otimes \boldsymbol{n}^{(i)}) \tag{6}$$

where *M* is the total number of active slip systems (up to 32 for the body centred tetragonal (BCT) crystal of β-Sn (Zhou et al., 2009) in SAC305), and $\boldsymbol{s}^{(i)}$ and $\boldsymbol{n}^{(i)}$ the slip and normal direction vectors of the *i*th slip system, respectively. The slip rate $\dot{\gamma}^{(i)}$ along the $i^{th}$ slip system is determined by temperature-dependent thermal activation events where pinned dislocations jump over obstacles (such as lattice defects, inclusions etc) in both forward and backward directions, which is given by:

$$\dot{\gamma}^{(i)} = \rho_{ssdm} b^2 v_D \exp\left(-\frac{\Delta H}{k\theta}\right) \sinh\left(\frac{\Delta V}{k\theta}\left|\tau^{(i)} - \tau_c^{(i)}\right|\right) \tag{7}$$

where ρ$_{ssdm}$ is the mobile statistically stored dislocation density, *b* the magnitude of the Burgers vector, *v$_D$* the dislocation jumping frequency, *ΔH* the activation energy, *ΔV* activation volume, *k* the Boltzmann constant, and $\tau^{(i)}$ and $\tau_c^{(i)}$ are the resolved shear stress and critical resolved shear stress on *i*th slip system, respectively. The critical resolved shear stress $\tau_c^{(i)}$ evolves with total dislocation density according to a hardening law.

A length scale dislocation density-based hardening law is adopted for the evolution of slip strength $\tau_c^{(i)}$ along the *i*th slip system:

$$\tau_c^{(i)} = \tau_{c0} + Gb\sqrt{\rho_{ssd} + \rho_{gnd}} \tag{8}$$

where $\tau_{c0}$ is the intrinsic critical slip strength, *G* the shear modulus, and $\rho_{ssd}$ and $\rho_{gnd}$ are the density of statistically stored (SSD) and geometry necessary dislocations (GND), respectively. The instantaneous rate of SSD density is directly determined by the current effective plastic strain *p* and its rate $\dot{p}$ in a recovery rule (Mecking and Kocks, 1981), where the recovery reflects the observation that slip resistance saturates when a large strain is applied (Lu et al., 2019; Zepeda-Ruiz et al., 2017) :

$$\dot{\rho}_{ssd} = \lambda_1 \dot{p} - \lambda_2 p \tag{9}$$

Here, $\lambda_1$ and $\lambda_2$ control the slip system hardening and recovery rates respectively. The effective plastic strain *p* is given by:

$$p = \sqrt{\frac{2}{3} \varepsilon^p : \varepsilon^p} \tag{10}$$

The GND density term $\rho_{gnd}$ in eq.(8) is calculated based on the plastic strain gradient tensor that accommodates lattice curvature (Arsenlis and Parks, 1999) after straining. Three components of the GND tensor, namely the screw dislocation density, the tangent edge dislocation density and the norm dislocation density, are found from a singular decomposition of the Nye's tensor, whose details can be referred to the local methodology in (Xu, 2021). A scalar GND density is then given by the by L$_2$-norm minimization of the three density terms (Cheng and Ghosh, 2015) as:

$$\rho_{gnd} = (\sum_{i=1}^{32} (\rho_{gs}^i)^2 + (\rho_{get}^i)^2 + (\rho_{gen}^i)^2)^{1/2} \tag{11}$$

GND density has been strongly correlated to the development of lattice curvature (Zhang et al., 2014) and intragranular mis-orientation (Witzen et al., 2020), where strain gradient is present. A numerical methodology is adopted to find the misorientation based on the resulting GND density within the investigated single crystal solders under thermal cycles, which is then compared to the experimentally measured misorientation from EBSD maps of solder cross sections. A brief description of the methodology is given below, with full details to be found in (Pantleon, 2008).

Lattice curvature tensor $\boldsymbol{K}$, as the reciprocal of curvature tensor $\boldsymbol{R}$, is related to the dislocation density vector $\boldsymbol{\alpha}$ and Burger's vector $\boldsymbol{b}$:

$$\boldsymbol{K} = 1/\boldsymbol{R} = \boldsymbol{\alpha}\boldsymbol{b} \tag{12}$$

From the geometry compatibility requirement, the lattice curvature tensor is also given by the gradient of lattice rotation:

$$\boldsymbol{K} = \frac{d\boldsymbol{\theta}}{d\boldsymbol{x}} \tag{13}$$

Therefore, the local misorientation with respect to a region with limited lattice rotation can be numerically found based on the GND density as:

$$\Delta\theta = \rho_{gnd} b \Delta x \tag{14}$$

where $\Delta x$ is the local mesh size.

A microstructure-sensitive, Griffith-Stroh type energy-density driving force is employed to investigate fatigue damage. The stored energy density originally proposed in (Wan et al., 2014) reflects the necessary requirement for crystalline slip, dislocation pile-up and interaction over a length scale with which the energy stored within dislocation structure is associated, and has been successfully applied to explain fatigue crack nucleation (Chen et al., 2018a) and the mechanisms of short crack growth in various alloys (Xu et al., 2021b). The stored energy density *G* is given by:

$$G = \frac{\xi U \Delta V_s}{\Delta A_s} = \int \frac{\xi \boldsymbol{\sigma} : d\boldsymbol{\varepsilon}^{\mathrm{p}}}{\sqrt{\rho_{ssd} + \rho_{gnd}}} \tag{15}$$

where *U* is the dissipated energy, $\Delta V_s$ and $\Delta A_s$ the dislocation storage volume and area over which the dislocation lines penetrate the surface, respectively, and $\xi$ is the fraction of plastic work stored as dislocation structures and estimated to be ~0.05 (Kamlah and Haupt, 1997). The stored energy in Eqn (15) depends explicitly on GND density, reflecting the contribution of local lattice curvature and crystal misorientation, and captures elastic free energy stored in dislocation interaction and structures. It provides a mechanistic driver of fatigue crack nucleation, and is utilised in this work, along with direct calculation of misorientation, to investigate crack nucleation of SAC305 single crystal solders under thermal cycling.

### 3.2.3 SAC305 material properties

The temperature dependent crystal plasticity material properties for SAC305 are obtained from a multi-scale homogenization model, which is established by virtue of integrating the mechanical behaviour of individual phases of β-Sn, $Ag_3Sn$ and $Cu_6Sn_5$ and their interactions (Xu et al., 2021a). The rate-dependent anisotropic material properties of β-Sn at room temperature are obtained from micro-pillar compression tests, while the properties of the intermetallics are based on the literature (Lucas et al., 2003). The temperature dependence is taken to be that observed in tensile experiments (Xie et al., 2021), and the temperature and strain rate-dependent stress strain relationship of a representative polycrystal has been developed from knowledge of micro-pillar single-crystal tests on the Sn phase and shown to reproduce the experimental results (Xu et al., 2021a). The model for SAC305 crystals is established based on the Sn phase properties and the IMC content and size effect through corresponding CPFE modelling in which both the pure Sn and the IMC phases are explicitly modelled in order to reproduce observed experimental stress-strain responses (Xu et al., 2021a). The material properties for the homogenized SAC305 single crystal are given in Table 3.

Table 3. Material properties for SAC305 single crystal at room temperature 25°C

| $E_{11}$(GPa) | $E_{33}$(GPa) | $G_{12}$(GPa) | $G_{13}$(GPa) | $\tau_{c0}^{(2,3)}$(MPa) | $\tau_{c0}^{(8,10)}$(MPa) |
|---|---|---|---|---|---|
| 26 | 70 | 24 | 22 | 4.8 | 7.5 |
| $\lambda_1(\mu m^{-2})$ | $\lambda_2(\mu m^{-2} s^{-1})$ | $\rho_{ssdm}(\mu m^{-2})$ | $k(\text{JK}^{-1})$ | $\Delta H$ | $\Delta V$ |
| 750 | 100 | 1.2 | $1.38 \times 10^{-23}$ | 41kJ/0.42eV | $9.26b^3$ |

# 4 Results
## 4.1 Extraction of solder displacements from continuum modelling

The thermo-mechanical simulation using the macro-model of the entire BGA assembly provides spatial and temporal predictions for the deformation, strain and stress responses of the entire BGA structure from which, in particular, the displacements imposed at the solder interfaces with the pads can be extracted. The solder material is modelled at this scale as non-linear visco-plastic, as discussed above, such that the analysis provides the inelastic strain energy density (plastic work) at the solder joints calculated in the continuum sense. The plastic work accumulated over a cycle with a stabilised hysteresis loop has been used as a damage metric for solders (Darveaux, 2002). During thermal cycling, the two temperature extremes (0°C and 100°C) define the displacement excursion limits of the assembly and specifically those at the solder joints. Figure 6 shows the contour plots of the global BGA in-plane (Z-direction) and out-of-plane (Y-direction) displacements at the two temperature extremes of the cycle. The continuum model assumptions for stress-free temperature explain the higher magnitude of deformation and higher residual stresses predicted at the low temperature extreme of the cycle.

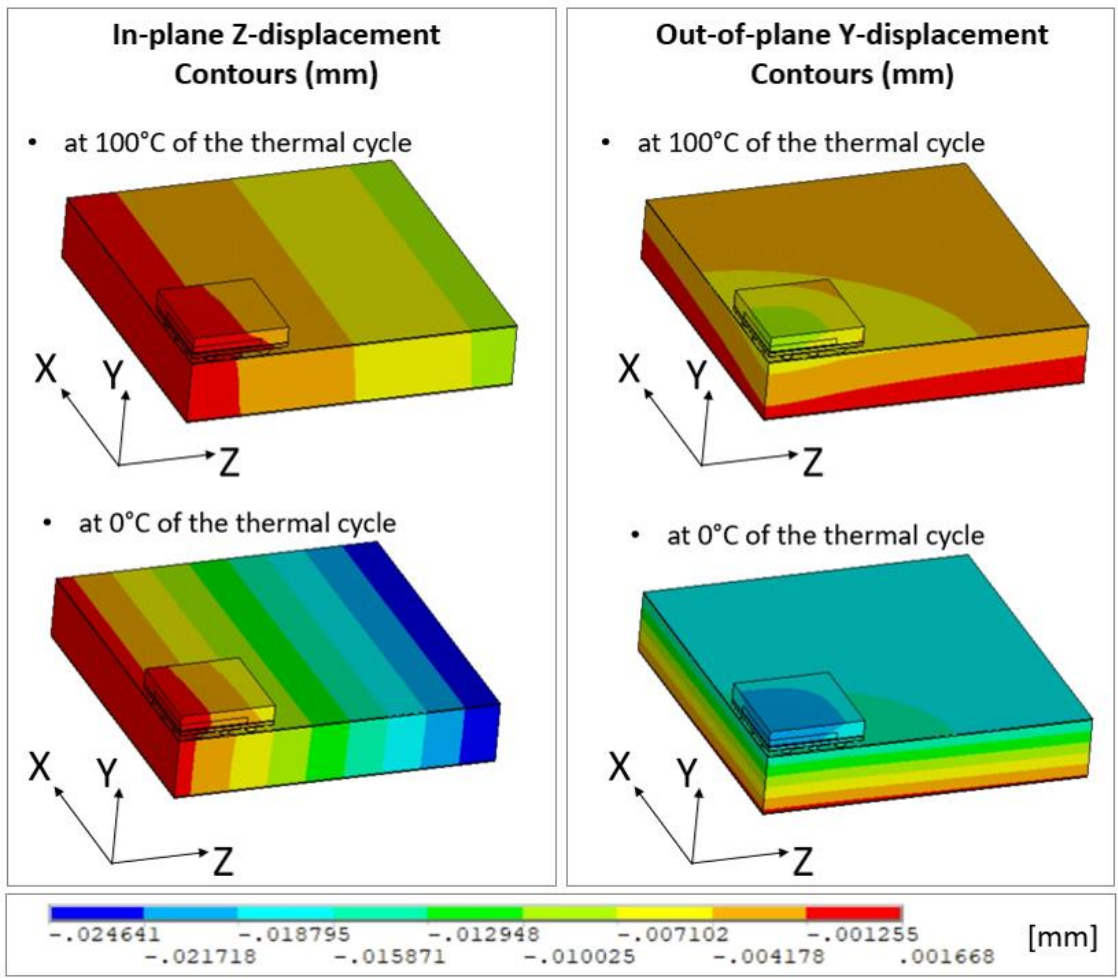

Figure 6. BGA assembly macro-model predictions for Z (in-plane) and Y (out-of-plane) displacement (mm) at the high (100°C) and low (0°C) temperature extremes of the thermal cycles.

The global deformations are determined by the structural loading and boundary conditions of the BGA continuum model, and the displacements of any given single solder under the thermal load may be quantified by the relative displacement field. This is obtained by processing the displacement nodal data of a given solder joint into a relative displacement field with reference to the node located at the bottom of the joint central line (in Figure 7(a), this node is labelled as N1). Figure 7 (a) shows an example of the contour levels of relative displacement in a solder joint at the low temperature extreme of the simulated load cycle #5. Figure 7 (b) is the related example of a relative node-to-node (N1-N2 nodes in Figure 7 (a)) movement under the simulated transient thermal load. The load profile includes the cooling down from the solder joint liquidus temperature in reflow to room temperature (resulting in residual stresses) followed by five consecutive thermal cycles which provide a stabilised hysteresis loop for the solder damage (fatigue and creep) estimation. This relative displacement field result, obtained from the BGA macro-model, is for the solder joint M6 in the grid array, but the same post-processing procedure of results has been applied in the same manner for all joints investigated in the study.

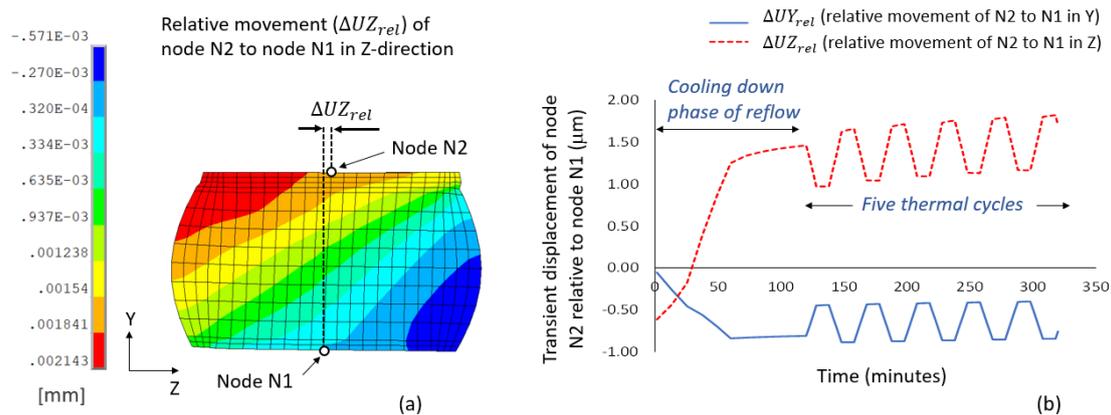

Figure 7. Macro-model displacement results for the M6 solder joint during thermal cycling: (a) Contour plot of the in-plane (Z) relative displacement field (relative to node N1 displacement) of the solder joint at the low temperature extreme (0°C), and the definition of the relative movement of node N2 to node N1 (deformed shape magnified by factor 5) , and (b) transient result for the in-plane (Z) and out-of-plane (Y) movement of node N2 relative to node N1.

The result for the relative (Y and Z) displacement of node N2 with respect to node N1 in Figure 7(b) confirm that the thermal cycles give rise to a progressive, permanent residual deformation of the solder joint, notably in-plane, driven by the thermal fatigue and creep damage mechanisms. However, the cyclic relative displacements can be seen to stabilise but with the superposition of a cyclic ratcheting displacement.

## 4.2 CP microstructural modelling of the solders

The y-component of the displacement field, which is passed from the continuum model, along the bottom face is shown in Figure 8(a) – (c) for Solder M6, M8 and M9 at the beginning of the 6$^{th}$ thermal cycle. The white arrow denotes the vector from the solder centre to the neutral point of the BGA board, and the Solder M9 is located more deviated, compared to M6 and M8, from the neutral point. The neutral point vectors are shown in Figure 8(d) for the three investigated solders from top view, and the deviation angles between the neutral point vector and the z-axis are tabulated.

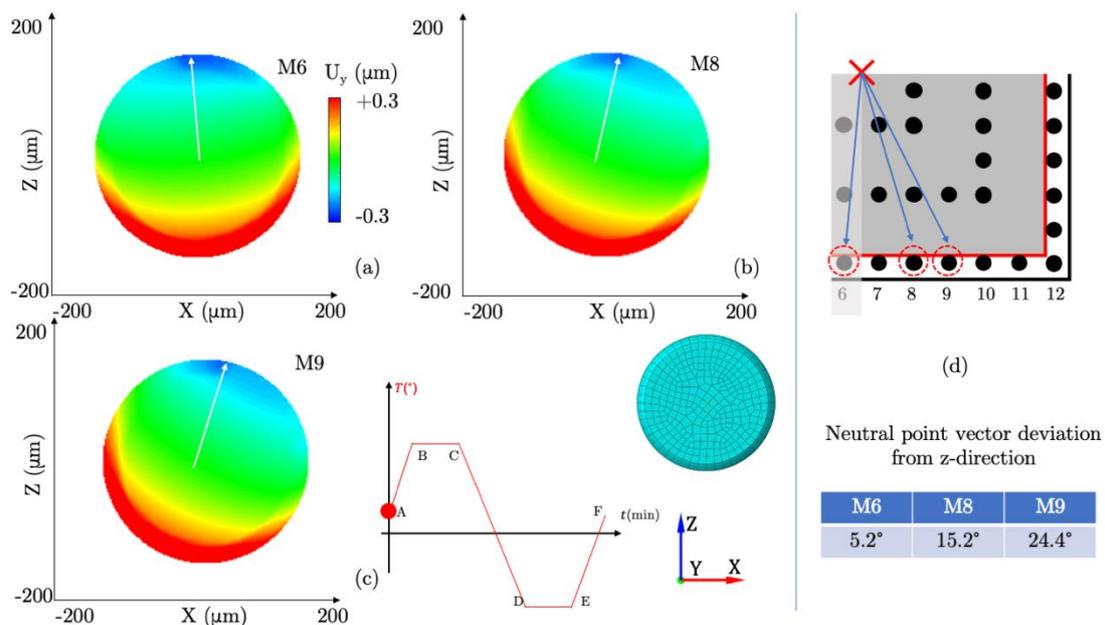

Figure 8. the y-displacement along the bottom face of the three solders M6, M8 and M9 at the first time point of Cycle 6, (a)-(c). The displacement field is obtained at the broad-scale model and pass to the individual solder-scale model. The locations of the three solders are also sketched in (d) with respect to the neutral point of the investigated BGA board.

The solder interfacial cyclic displacements so obtained from the BGA-scale model are then imposed on explicit crystal plasticity model representations of solder joints M6,

M8 and M9 as shown in Figure 1. The slip system activations contribute to the accumulative plastic strain (given by equation 10) and this quantity calculated from the CP solder modelling for the three differing solder joints is shown in Figure 9(a) at the end of $6^{th}$ and $10^{th}$ cycles. The plastic strain trend is in line with the inelastic energy predicted by the continuum-scale modelling as shown in Figure 7. The plastic strain along a circumference (A-A') of the solder located on the upper interface is shown along with contour plots for the three joints considered. Plastic strain has often been considered an important precursor to microstructure-sensitive crack nucleation (Prastiti et al., 2020; Sangid, 2013), though is argued strongly to be a necessary as opposed to sufficient driver, and this also applies to solder alloys (Lee et al., 2002). The highest strains develop at the top of the solders, resulting from the constraint imposed because of the mismatch in thermal expansivity between the BGA package and the solder. The solder joints M6, M8 and M9 will have differing microstructure (including $\beta$-Sn crystallographic orientations) as well as differing loading/constraint imposed upon them by the BGA as a result of their differing positions. These differences are anticipated to lead to different strain histories and this is observed in Figure 9(a), with solder M9 being disadvantageously loaded to give the highest plastic strain accumulation of the three considered, and solder M6 the lowest. This is consistent with the experimental observation that solder M9 is the most damaged (see Figure 2). For each solder, however, a heterogeneous distribution of plastic stain develops which in general will be very dependent on solder crystallographic orientation (see later). The cyclic stored energy density rate ($\dot{G}$), calculated as the average over cycles six to ten, which is a mechanistic driver of microstructurally-sensitive fatigue crack nucleation is shown in Figure 9(b). This quantity overall suggests the same anticipated ranking on fatigue life as the plastic strain. The solder locations within the array and their crystallographic orientations differ, both of which will influence fatigue damage during thermal cycles, and are explored next.

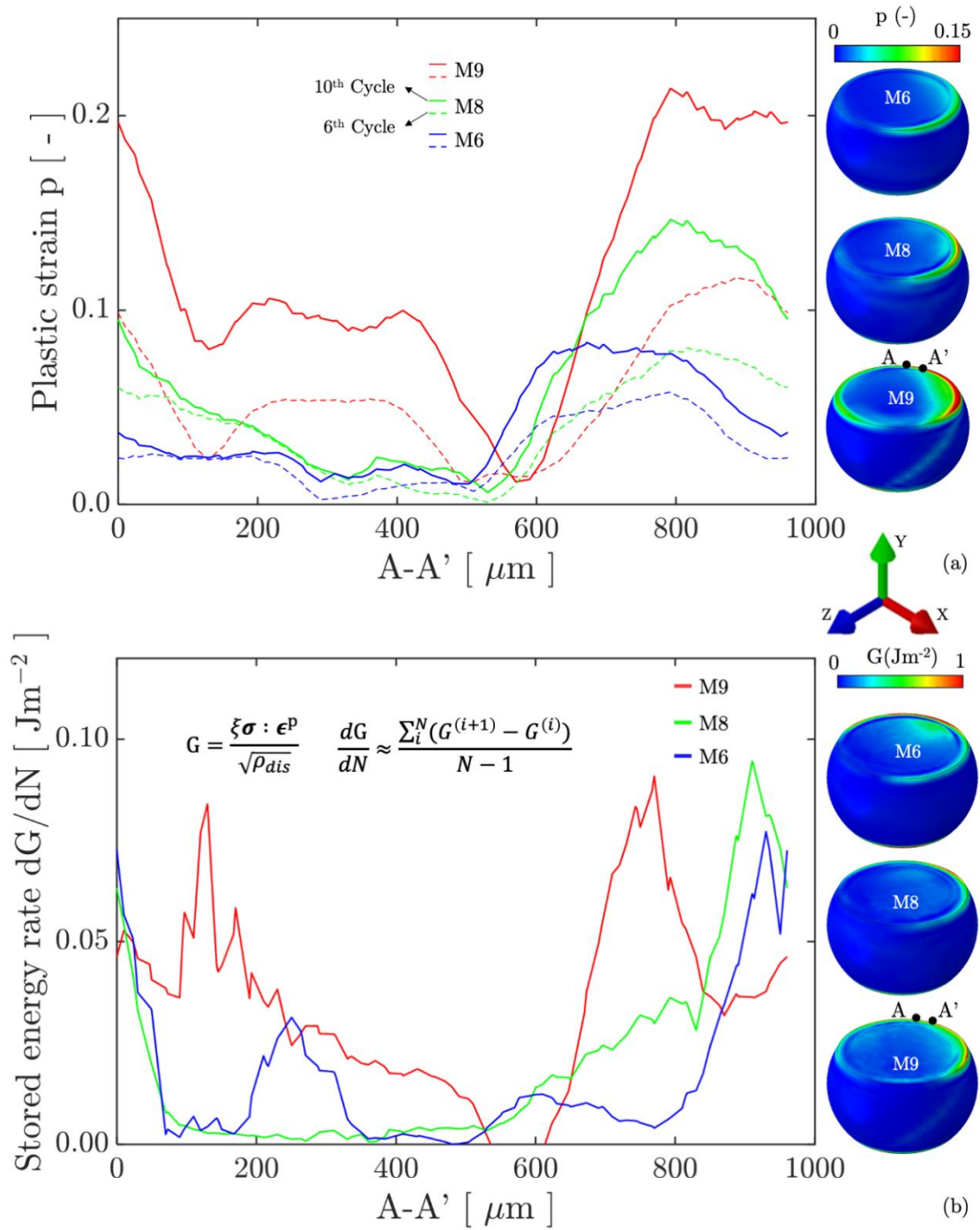

Figure 9. (a) Plastic strain accumulation at cycles 6 and 10, and (b) stored energy density rate at cycle 10 for solders M6, M8 and M9.

Geometrically necessary dislocations (GNDs) are those which support crystal lattice curvature (in the absence of remote stress) and can be calculated from plastic strain gradients. The magnitude of the GND density is therefore a simple measure of the magnitude of local lattice curvature, or intragranular misorientation within single crystals, which has been shown to correlate to micro-crack nucleation (Sweeney et al., 2013) and recrystallization (Chen et al., 2020). The CP model facilitates the calculation of GND density and for each solder joint, a vertical mid-plane section is chosen to show their distributions within solder M6, M8 and M9 in Figure 10(a). Highly localized GND densities are found to develop near the top interface of the three solders because of the strong local plastic strain gradient (Arsenlis and Parks, 1999) induced by the thermal loading (and shown in Figure 10(a)). The highest densities of GNDs are predicted to occur within solder M9 followed in turn by M8 and M6. The presence of GNDs is a significant contributor to the stored energy density since it reflects local dislocation interactions and pile-ups (and hence configurational energy) as well as the maintenance of lattice curvature, and is indicative of microstructural damage, particularly fatigue crack nucleation, suggesting potential locations for referred nucleation (Sangid et al., 2011). A quantitative comparison of the CP calculated (using Eqn.(14)) and experimentally measured intragranular misorientation is possible, which is calculated with respect to the centre region of the cross-section (indicated as the white frame), shown in Figure 10(b) and (c). The ranking of misorientation magnitude is captured reasonably well by the CP calculated misorientation for the three solders which as anticipated reflects the GND density results. The relatively low magnitude of the experimental measurements by comparison results from the fact that recrystallization had taken place when the measurements were made after at least 1,000 thermal cycles. At this point in the cycling, the stored energy is anticipated to have reduced considerably following energy release by the recrystallization process (Belyakov et al., 2002; Gourdet and Montheillet, 2003), while the numerical results present the mis-orientation prior to potential recrystallization and hence maintain a higher magnitude.

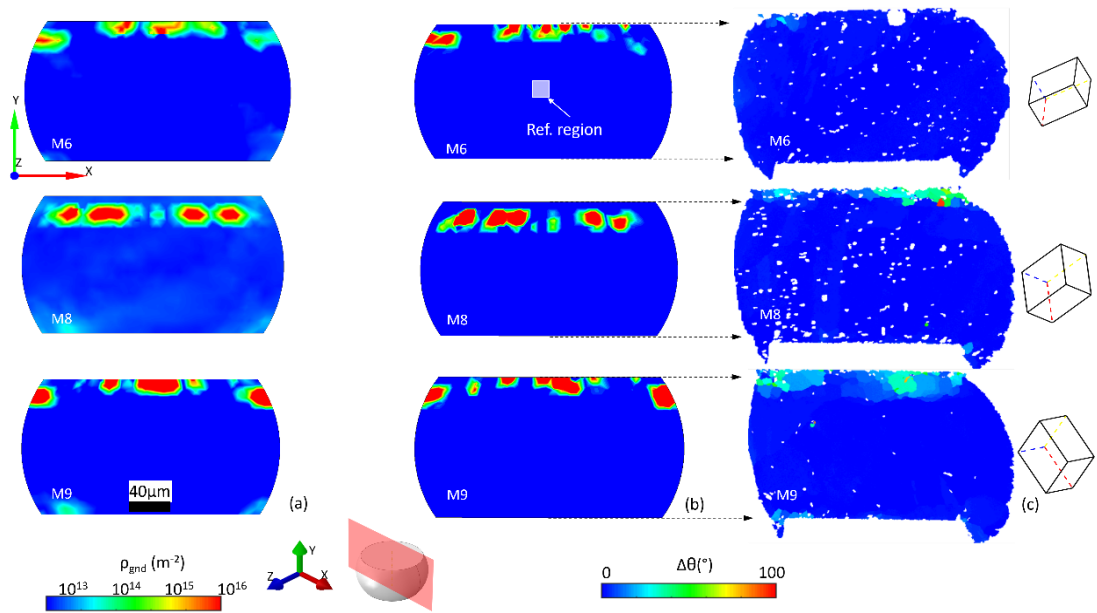

Figure 10. (a) the GND distribution along the z-cross section of the three single grain solder joints at the end of the 10$^{th}$ cycle. Crystal misorientation distributions (b) CP calculated, and (c) experimentally measured with EBSD along the Z-normal section of the three solders M6, M8 and M9

An analysis has been carried out of the potential relationship between the misorientation developed and the crystal c-axis angle (between the c-axis and the Y-axis, *i.e.* the substrate norm), for the three solders and is shown in Figure 11. The averaged misorientation within the cross-section selected, in both the CP model and experimental results, is found to be strongly dependent on the c-axis angle with respect to the substrate plane. A solder with its c-axis close to parallel to the BGA board (substrate) is found from measurement to incur more damage (in terms of misorientation), and this relationship is also similarly predicted by the CP model. In addition, a consequence of the high misorientation is the CP calculated stored energy density reflecting predicted fatigue life. Hence fatigue crack nucleation (Wang et al., 2021) and recrystallization (Belyakov et al., 2002) as indicated in Figure 10(c) are likely to develop at the solder upper interface because this is the location at which misorientation, and hence GND density and in turn stored energy density, are highest. The misorientation may be determined by averaging over the entire section area considered, or by selecting the top partition of the whole in which most of the misorientation develops (here defined as the 40μm depth from the upper interface). Both measures are included in Figure 11. The local measurement suggests a significantly higher value of misorientation compared to that measured within the whole cross-section. This further supports that crack nucleation (and recrystallisation) is caused by the localization of misorientation, arising from the strain gradients which develop near the upper interface. Though located at three different locations on the BGA board, the joint location effect is shown to be minimal (see Figure A 1 in Appendix) compared to the crystallography effect, because locations M6, M8 and M9 are adjacent to each other.

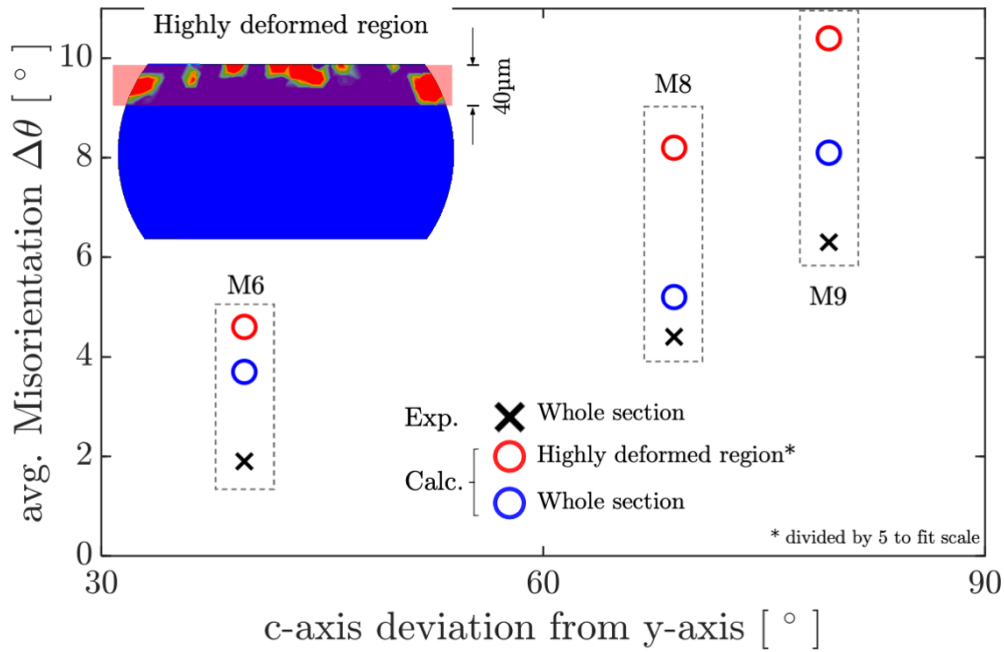

Figure 11. Relationship between the averaged misorientation and the crystal c-axis direction for the three single grain solders. (The Y-axis lies perpendicular to the substrate.)

The misorientations developed and the cracks nucleated within single crystal solders M8 and M9 are shown in Figure 12 after thermal fatigue cycles. Highly localized misorientation and crack nucleation sites are observed near the top interfaces to the pad, which arises from the thermal mismatch between the Sn-based solder and the copper substrate in addition to the mechanical constraint. The consistency and strong correlation of the sites (indicated by the red arrows) between the misorientation hotspots and crack nucleation indicates that the GND density carrying the lattice curvature during plastic deformation, which is directly correlated to the misorientation (see Figure 11), is potentially capable of capturing the fatigue crack nucleation and damage within solders. The stored energy, which is strongly influenced by the GND density and has been successful in predicting fatigue crack nucleation in various alloys (*e.g.* (Chen et al., 2018a)), is assessed against these experimental observations next.

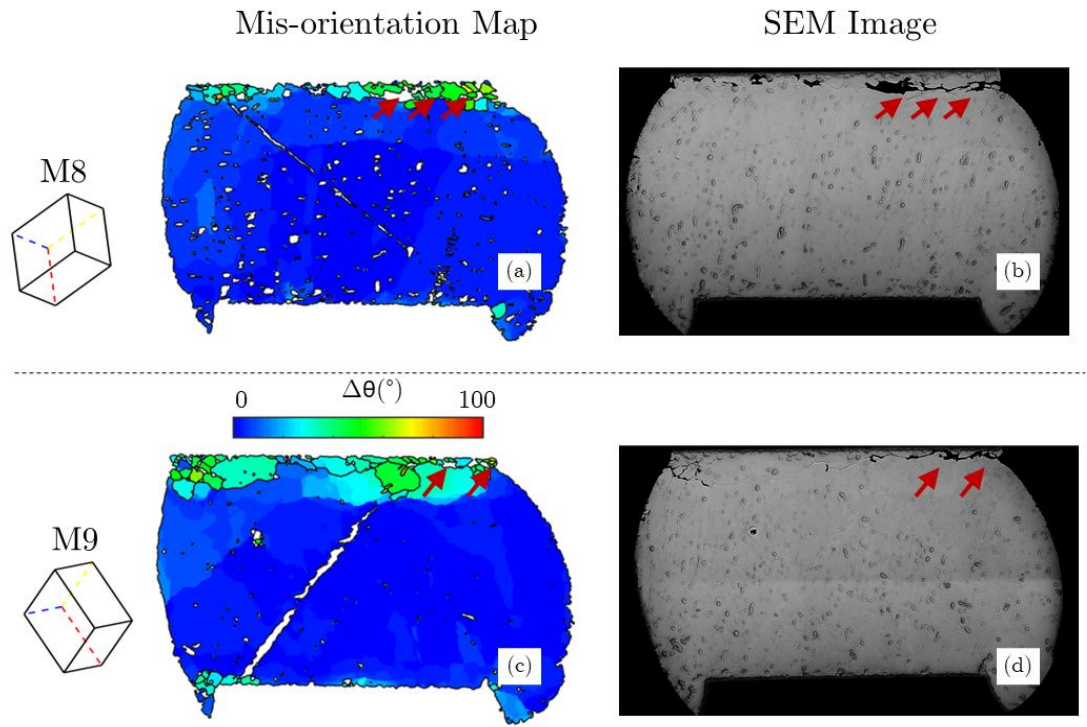

Figure 12. The misorientation development ((a) and (c)) and crack nucleation ((b) and (d)) in solder M8 and M9, respectively, during thermal cycles, which are characterized by using EBSD and SEM.

4.3  Tin crystallography effects

The misorientation results above suggest a strong relationship between the solder crystallographic orientation (given by its c-axis) and microstructural damage, which could eventually lead to fatigue crack nucleation. Hence in this section, we present a systematic study of orientation effects on plastic strain and cyclic stored energy density using the solder CP model. The crystallographic effect is first investigated by considering two extreme c-axis orientations, namely perfectly parallel and perpendicular to the substrate, corresponding to the displacement conditions for solder M9. The results for the actual (measured) orientation of solder M9 (as in the section above) are also included. The resulting CP calculated plastic strain distributions along the top circumference A-A' of the solder are shown in Figure 13(a) for the three cases. The solder with c-axis parallel to the substrate develops the highest magnitude of plasticity, which is followed by the experimentally measured M9 orientation. The microstructure with c-axis perpendicular to the substrate strongly suppresses the plasticity developed within the solder, which benefits the fatigue lifetime of the solder. This optimum crystallographic orientation arises because of anisotropy of thermal expansion, crystal elasticity and plasticity as the [001] direction has the highest elastic modulus (Rayne and Chandrasekhar, 1960) and the slip systems are difficult to activate (Kaira et al., 2016). The cyclic stored energy rates are reported in Figure 13(b) for the three cases. Correspondingly, the stored energy for c-axis parallel to the substrate is higher and as much as three times that for the c-axis perpendicular to the substrate. This suggests considerably longer life to crack nucleation for the latter crystallographic orientation according to mechanistic studies of crack nucleation (Chen et al., 2018a) and growth (Wilson and Dunne, 2019). However, experimental validation of stored energy density along the section surfaces of solders remains to be obtained, but a methodology for this has been established with digital image correlation (DIC), as reported in (Xu et al., 2021b). The effects of c-axis orientation observed and predicted with CP modelling are consistent with independent experimental observations (Bieler et al., 2008).

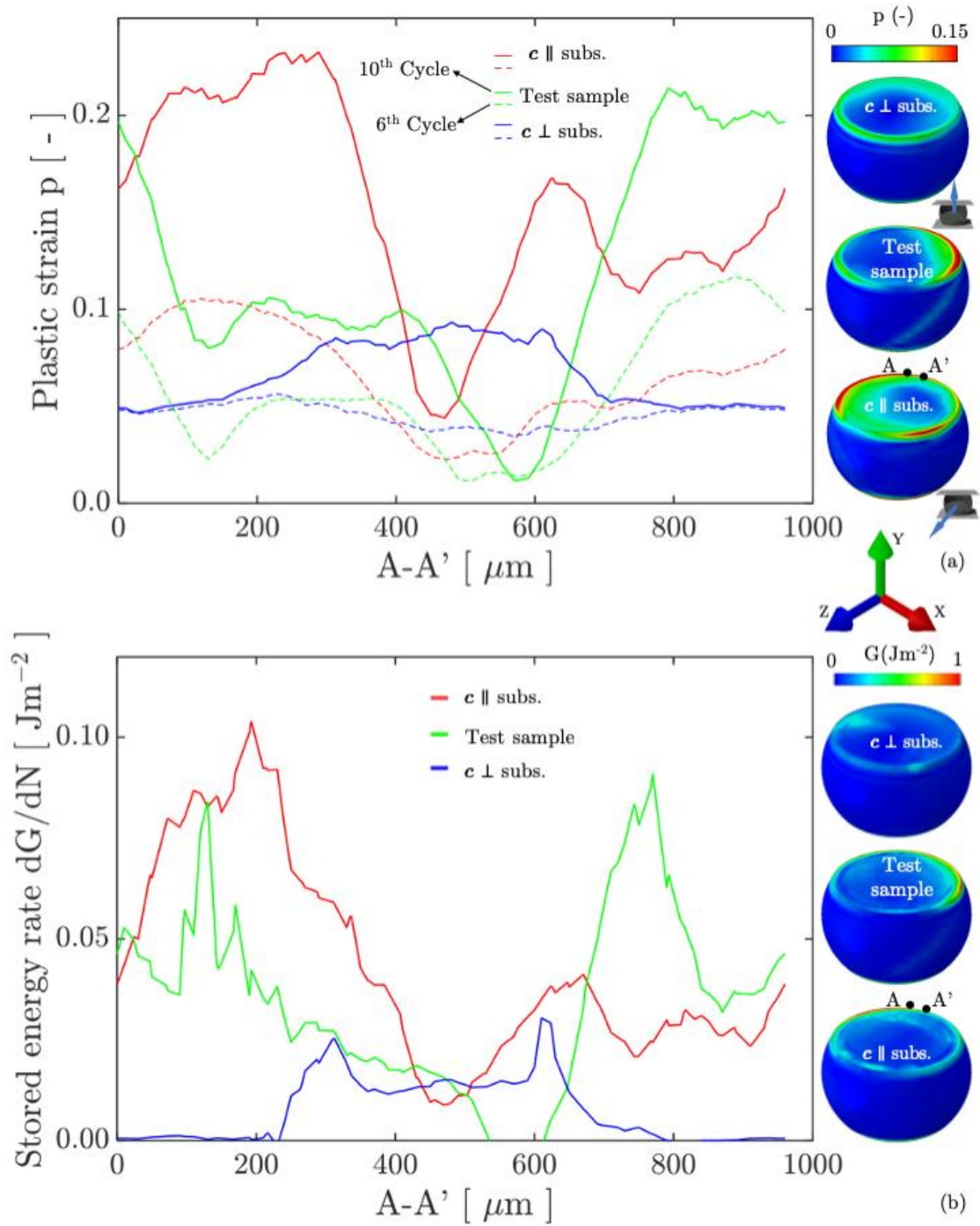

Figure 13. the effect of c-axis orientation on (a) plastic strain and (b) cyclic stored energy along Path A-A' of Solder M9 with three crystallographic orientations. The contour results are shown at the end of the 10th cycle. In (a), the dash lines refer to plastic strains at the end of 6th cycle while the solid lines refer to those at the end of 10th cycle.

In addition to the c-axis, the effect of the crystal a-axis orientation is also investigated by fixing the c-axis direction and studying the role of a-axis orientation, again by considering joint M9. The crystal orientations for three realizations plus that experimentally measured are illustrated in Figure 14(a). The plastic strain and cyclic stored energy density contours are shown for the four orientations in Figure 14(b). The distributions are changed much less than for changing the c-axis orientation (Figure 13(a)) because of the crystal anisotropy, with small differences in magnitude developing, which are detailed in the profile plots in Figure 14(c) and (d). The resulting insensitivity arises because the anisotropy of the crystal within the (001) plane is low.

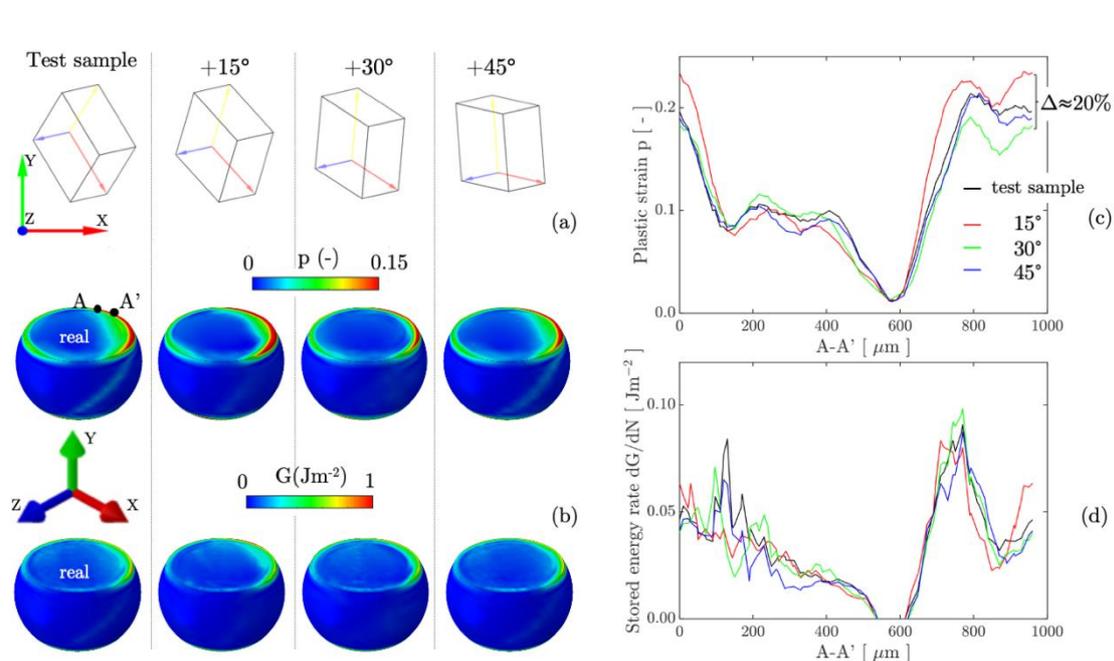

Figure 14. The effect of a-axis orientation on damage development in solder M9. (a) Unit cells for the four a-axis orientations. (b) Plastic strain and stored energy distribution within the solders at the end of the 10$^{th}$ cycle, and the distributions of (c) plastic strain, and (d) cyclic stored energy density along Path A-A'.

With the establishment of the relationship between the solder damage arising with c-axis orientation to the substrate, an additional study is carried out, in which the single crystal is rotated about the Y-axis for understanding the c-axis deviation to the substrate's norm. The unit cells for five representative crystal realizations plus the experimentally measured crystal orientation are illustrated in Figure 15(a), and the corresponding plastic strain and cyclic stored energy density contours are shown in Figure 15(b). The distributions of both quantities are clearly affected by the angular position of the crystal about the c-axis when aligned with the Y-axis, which is further detailed in the line plots along the A-A' circumference in Figure 15(c) and (d). The single crystal orientation experimentally measured and that for +45° introduces more damage to the solder compared to other orientations.

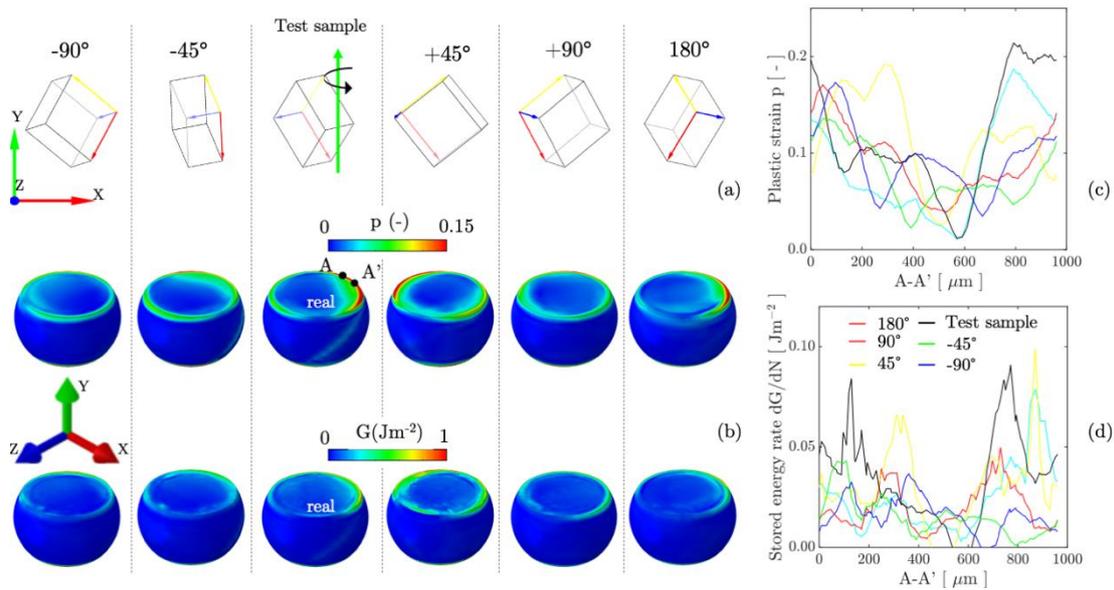

Figure 15. The effect of crystal orientation when rotated about the normal to the BGA substrate (i.e. the Y-axis indicated in a green arrow shown in the experimental measured unit cell). (a) Unit cells for the 6 crystallographic orientations considered, and (b) plastic strain and cyclic stored energy at the end of the 10$^{th}$ cycle. The distributions of (c) plastic strain and (d) cyclic stored energy rate along Path A-A'.

A deviation angle ω, which is defined as the angle between the neutral axis vector (see Figure 8(d) for the definition) and the projected c-axis onto the substrate, is correlated to the CP calculated plastic strain and cyclic stored energy rate for solders with orientations given in Figure 15. The dependence of plastic strain and stored energy density on the deviation angle ω is shown in Figure 16. The damage, in terms of plastic strain and cyclic stored energy, seems to be increased as the two vector directions converge. The neutral axis vector represents the maximum shear stress direction (Che and Pang, 2013), and solder slip systems are more favourable to be activated when the c-axis approaches to the maximum shear direction (Zhou et al., 2015). Hence plastic strain and stored energy are likely to be higher. Therefore, a large deviation ω may alleviate the resulting fatigue damage. Of more importance, this finding may provide indicative formulae for providing bespoke optimum individual solder orientations even on the same BGA board, as the neutral axis vectors vary with joint location in the array.

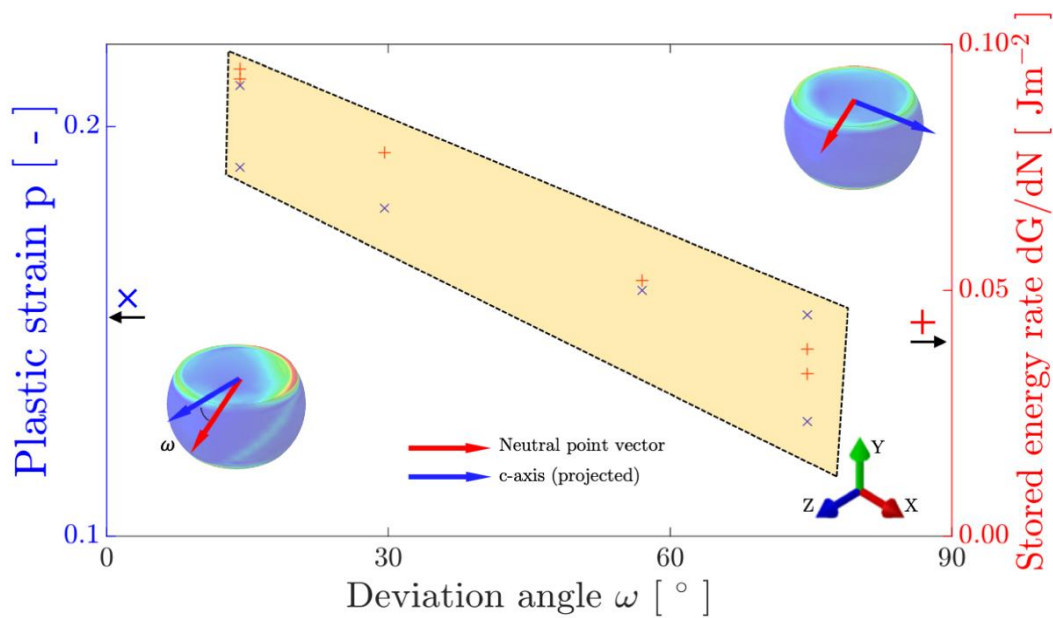

Figure 16. Relationship between solder damage (plastic strain and cyclic stored energy density) and deviation angle $\omega$ between the c-axis and neutral axis vectors.

Multiscale modelling has been established which links from solder-board thermo-mechanical conditions through to local behaviour of individual solders and their unique microstructures. This allows the possibility of bespoke microstructure solder design for particular array locations as well as thermo-mechanical loading conditions.

Page 38 of 47

# 5 Conclusions

An integrated multi-scale modelling methodology with quantitative experimental characterisation has been developed to investigate the mechanistic basis of damage within SAC305 single crystal solder joints subject to thermal cycling. The key role of β-Sn crystal orientation in single crystal joints has been comprehensively investigated. The model is also well suited to be extended in the future to include different initial IMC particle sizes, and multi-grain joints with morphologies such as beachball and interlaced.

- The microstructural model, where the boundary conditions were taken from the continuum model, has shown that the damage within solders is highly correlated to the plastic strain inhomogeneity and misorientation developed in the β-Sn near the top (package-side) interface, in agreement with experimental measurement.

- The microstructure, where the c-axis ([001]) is perpendicular to the substrate, is predicted to be the most resilient to thermal fatigue while the one with c-axis parallel the most vulnerable. Compared to the determining effect of the c-axis, the a-axis shows limited effect. In addition, the damage is affected by the deviation angle between the c-axis and the neutral axis vector of solders with respect to the package centre. The β-Sn orientation effect plays a primary role in determining the damage of single crystal joints under thermal fatigue.

- The underpinning mechanisms of the orientation effect on solder damage has been rationalized as the anisotropy of thermal expansion, elasticity and plasticity of SAC305 alloy. This study suggests the optimum orientation of solder joints to be fabricated for thermal fatigue resistance.

# Acknowledgement

All authors acknowledge the financial support by the Engineering and Physical Sciences Research Council for funding through the grants EP/R018863/1 and EP/R019207/1. FPED wishes to acknowledge gratefully the provision of Royal Academy of Engineering/Rolls-Royce research chair funding.

# Appendix

The location effect is investigated by assigning the boundary conditions that were obtained from the board-scale model for Solder M6, M8 and M9 but their crystallographic orientations are set to be the same. The plastic strain developed at the 6$^{th}$ and 10$^{th}$ cycle within the three solders is shown in Figure A 1 (a) and (b) for the orientation as c-axis parallel (a) and perpendicular (b) with respect to the substrate, respectively. For the microstructure when the c-axis is parallel to the substrate, the profile of plastic strain along the top surface shows more inhomogeneous distribution when solders are placed farther away from the centre point, which is given by the strong asymmetry of the displacement fields (see Figure 8). This indicates that the solder sitting the corner of the BGA board (*e.g.* M12 Figure 1) would probably fail first provided the microstructures of all the solders are fabricated close to this microstructure. However, the location effect substantially diminishes when the c-axis is perpendicular to the substrate (b) and the cycle number is sufficiently large. The location effect on the damage of solders is hence coupled to the detailed microstructures of them, and the microstructure effect outplays the location effect in terms of damage, at least for the solder locations investigated herein.

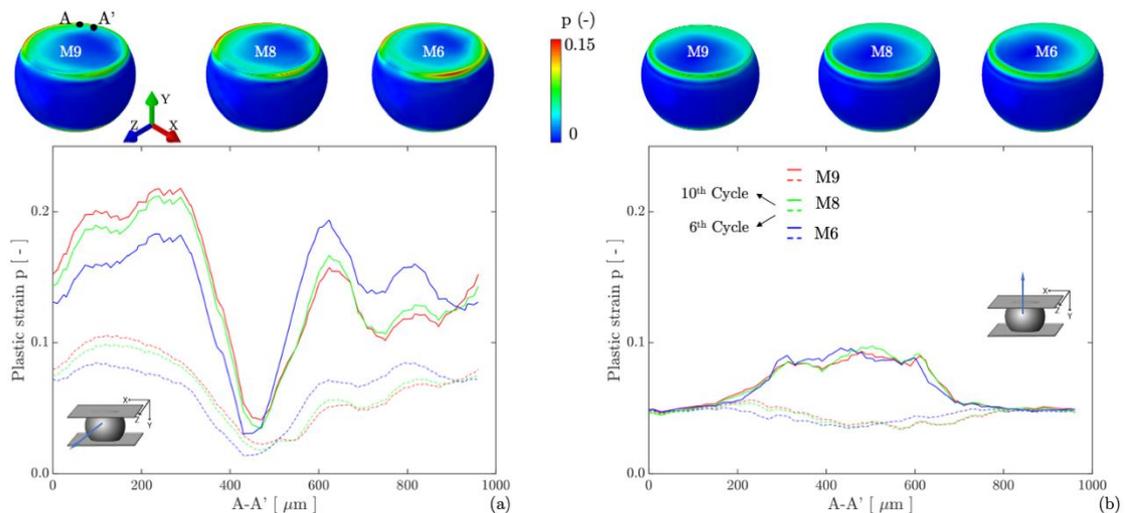

Figure A 1. the plastic strain developed within three solders M6, M8 and M9 subject to the boundary conditions passed from the global model but with same microstructure (a) when the c-axis is parallel and (b) perpendicular to the substrate, respectively.

Zhou, B., Bieler, T.R., Lee, T.K., Liu, K.C., 2009. Methodology for Analyzing Slip Behavior in Ball Grid Array Lead-Free Solder Joints After Simple Shear. Journal of Electronic Materials 38, 2702-2711 DOI: http://doi.org/10.1007/s11664-009-0929-6.

Zhou, B.T., Zhou, Q., Bieler, T.R., Lee, T.K., 2015. Slip, Crystal Orientation, and Damage Evolution During Thermal Cycling in High-Strain Wafer-Level Chip-Scale Packages. Journal of Electronic Materials 44, 895-908 DOI: http://doi.org/10.1007/s11664-014-3572-9.